%% file: main.tex
\documentclass[sigconf]{acmart}

\usepackage{hyperref}
\usepackage{xcolor}
\usepackage{graphicx}

\usepackage[utf8]{inputenc} 
\usepackage[T1]{fontenc}    
\usepackage{makecell}
\usepackage{natbib}
\usepackage{subcaption}
\usepackage{pgf-pie} 
\usepackage{pgfplots, pgfplotstable}
\usepackage{tikz}
\usetikzlibrary{shapes,backgrounds}
\usepackage{booktabs}
\usepackage{fontawesome5}
\usepackage{tabularx}

\AtBeginDocument{%
  \providecommand\BibTeX{{%
    \normalfont B\kern-0.5em{\scshape i\kern-0.25em b}\kern-0.8em\TeX}}}

\setcopyright{acmcopyright}

\copyrightyear{2022}
\acmYear{2022}
\acmConference[WWW '22 Companion] {Companion Proceedings of the Web Conference 2022}{April 25--29, 2022}{Virtual Event, Lyon, France.}

\begin{document}

\title{
Personal Knowledge Graphs: Use Cases in e-learning Platforms}

\author[Eleni Ilkou]{Eleni Ilkou \\\small{supervised by Prof. Dr. Wolfgang Nejdl}}
\email{ilkou@l3s.de}
\orcid{0000-0002-4847-6177}
\affiliation{%
  \institution{L3S Research Center, Leibniz University Hannover}
  \streetaddress{Appelstr. 9}
  \city{Hanover}
  \country{Germany}
  \postcode{30167}
}

\renewcommand{\shortauthors}{Eleni Ilkou}

\begin{abstract}



Personal Knowledge Graphs (PKGs) are introduced by the semantic web community as small-sized user-centric knowledge graphs (KGs). PKGs fill the gap of personalised representation of user data and interests on the top of big, well-established encyclopedic KGs, such as DBpedia. Inspired by the widely recent usage of PKGs in the medical domain to represent patient data, this PhD proposal aims to adopt a similar technique in the educational domain in e-learning platforms by deploying PKGs to represent users and learners. We propose a novel PKG development that relies on ontology and interlinks to Linked Open Data. Hence, adding the dimension of personalisation and explainability in users' featured data while respecting privacy. 
This research design is developed in two use cases: a collaborative search learning platform and an e-learning platform. Our preliminary results show that e-learning platforms can get benefited from our approach by providing personalised recommendations and more user and group-specific data.

\end{abstract}



\maketitle

\input{1-introduction}

\input{2-problem}

\input{3-sota}
\input{4-approach}
\input{5-methodology}

\input{6-results}

\input{7-conclusion}

\section*{Acknowledgements}


The author would like to thank Prof. Dr. Maria-Esther Vidal for the fruitful discussion, guidance, and insightful comments. This work is funded by the European Union's Horizon 2020 research and innovation program under the Marie Sk\l{}odowska-Curie project Knowgraphs (grant agreement ID: \href{https://cordis.europa.eu/project/id/860801}{860801}).

\bibliographystyle{ACM-Reference-Format}
\bibliography{main}

\end{document}

%% file: 1-introduction.tex
\section{Introduction/Motivation}

Nowadays, societies focus on the digital transformation of education and skill development in online learning platforms~\cite{davies2020developing}. Millions of learners use daily online learning platforms for their formal education, especially during the COVID-19 pandemic. The need for online teaching and lifelong learning tools has gained momentum. 
The same happens with online collaborative learning and search. Collaborative search happens when two or more people team up in a search task online and perform synchronous or asynchronous searches. Collaborative work online and collaborative learning are more important than ever; however, platforms supporting web collaboration lack semantic features, such as interconnections between the data and semantic recommendations. These platforms are mainly customised to facilitate learning applications and usually ignore the description and documentation of modelled concepts. As a result, current approaches cannot exploit common understanding encoded either in domain ontologies or knowledge graphs.
At the same time, there is an increased need for 
systems with high personalised capabilities, personalised collaborative search~\cite{DBLP:journals/jms/CR19}, and more productive and impactful platforms that can support collaborative learning. Semantic web (SW) technologies can assist in this effort. 

Knowledge Graphs (KGs), although well-established and widely used, such as the DBpedia and Yago, do not directly contain personal information and are not usually designed to accommodate users' personal data. The same issue occurs also in domain KGs, such as in educational domain, where the need for personalisation is significant~\cite{ilkou2020technology}. The recently suggested personal knowledge graphs (PKGs)~\cite{DBLP:conf/icdm/SafaviBFMMK19} come to fill this gap by offering pocket-sized knowledge graphs related to users' interests. The same issue also occures in educational KGs where . PKGs in online learning environments can benefit users by providing personalised features 
and connecting their actions on the web with KGs and Linked Data. 

The proposed line of work aims to assist the efforts of the SW community in setting the standards of new applications with deploying further research in SW technologies, such as the PKGs, in no classical computer science domains of application; in our case, the educational domain and the collaborative learning and e-learning sub-domains. Also, based on our knowledge, this is the first attempt to utilise PKGs for purely educational and learning purposes and have them developed in e-learning platforms. This work can assist researchers in both fields of education and computer science to create better personalised systems, and encourage developers in e-learning platforms to utilise SW technologies.
This proposal aims to explore the syntax and semantics of PKGs in learning environments and increase the personalisation and explainability of the learning system's actions to users.






%% file: 2-problem.tex
\section{Problem}

The PhD thesis addresses the problem related to the personalisation of the web searches and 
to the enhancement of e-learning environments with personalised features linked to the SW. 
For example in e-learning platforms the importance of personalised recommendations is increasing; the PKGs could assist in this effort an e-learning platform which uses a knowledge base for its content.
Also, collaborative searching environments and collaborative learning usually offer features adopted from non-collaborative settings. Not many features are developed directly for collaborative learning environments, which creates a gap between the needs of collaborative environments and the actual technology offered. A solution could be the connection of 
the collaborative search with the Linked Data.  
This can occur by utilising PKGs to connect individuals' and group's search with the semantic web and offer a better collaborative and personalised experience. 

The thesis is formulated around the analysis of the characteristics related to the deployment of PKGs for educational applications. However, we can especially classify our use cases in collaborative learning environments and e-learning platforms. Therefore, we formulate the research questions of the thesis as follows:

\begin{itemize}
    \item RQ1: How to syntactically and semantically represent a PKG in the e-learning domain?
    \item RQ2: How can e-learning platforms offer better semantically enhanced personalised features, such as semantic recommendations, with the usage of PKGs? 
    \item RQ3: How can collaborative search learning environments offer more personalised features with the usage of PKGs? 
    \item RQ4: Can collaborative learning platforms offer better collaboration with the usage of PKGs? 
\end{itemize}

%% file: 3-sota.tex
\section{State of the art}

\subsection{Personal Knowledge Graphs}

Personal or Personalised Knowledge Graphs (PKGs) are small graphs on top of KGs which contain user's related data. They are the natural complementary effort to address the challenges raised by the personal information management (PIM)~\cite{DBLP:journals/arist/Jones07}, about users retrieving and organising personal information. 
PKGs are currently mostly used in the medical domain, on top of medical KGs to represent patients data~\cite{wang2018design}.

In the broad domain, the works of Personalised Knowledge Graph Summarization~\cite{faber2018adaptive, DBLP:conf/icdm/SafaviBFMMK19} were introduced. The PKG summarisation constructs personal summaries of users from a KG that contain the relevant facts of the users' interests. These summaries support personalised content queries and utility. These works are linked to KGs, such as the DBpedia and Yago, and provide the theoretical formulations for constructing PKG summaries from users' past queries. 
Moreover, similar work to the PKG summaries direction suggests graph-based approaches for users activity discovery from heterogeneous personal information collections, such as emails and files~\cite{DBLP:conf/wsdm/SafaviFSJWFKB20}. It proposes a method for unsupervised setting that perceives privacy. Influenced by these works, we intend to create 
PKGs containing information about users' and groups' queries and actions while paying attention to privacy.


\subsection{Personalisation in e-Learning}

Personalisation in e-Learning is about creating an adaptive environment for the learner with adaptive content that derives mainly from the curriculum, educational resources, learning path, learning preferences, and cognitive state. Then the personalisation usually happens with a recommender system that suggests relevant content to the user, in terms of content type and characteristics as well as previous actions and performance of the user. The goal is to create user-centric applications that aim towards a positive learning experience. For this objective, there are two main approaches, a symbolic, with the usage of ontologies, and a sub-symbolic approach, in which they handle recommender as a black box. 

The symbolic approaches are deployed with the usage of ontologies and semantic frameworks. WASPEC~\cite{apoki2021design} utilises a semantic framework that models learner profiles and categorises personalisation parameters to learning preferences and accessibility. Our latest publication, EduCOR ontology~\cite{DBLP:conf/semweb/IlkouATHKAN21}, contributes in that line as it is analysed in Section~\ref{sec: results}.
Another interesting approach is the formulation of the learning session graph to achieve personalisation in informal learning settings in information wikis~\cite{ismail2019framework}. Their framework achieves high scores in recommendation relevance. Encouraged by their design and findings, we target integrating session graphs in modelling user activity in the PKG.

An alternative method is a "black-box" educational recommender system can benefit from open learner models providing a better sense of learning~\cite{abdi2020complementing}. An open learner model assists students to understand their learning process and their peers. It can be classified as a cognitive tool approximating students' abilities with a score similar to learning analytics dashboards. Abdi et al.~\cite{abdi2020complementing} showed that open learner models could increase the explainability and transparency of a recommendation; however, there were concerns regarding the fairness and feedback of the recommendation. Their study suggests that users' input and feedback into the recommendation process can improve satisfaction. Motivated by their findings, we aim to implement a transparent model in PKGs which justifies system actions and considers users' input and feedback in the recommendation process while offering additional personalisation via semantic enrichment.



\subsection{Collaborative Search and Learning}

Collaborative search can occur as a learning activity; Searching as Learning (SaL) explores this direction and links collaborative search with collaborative learning. SaL in a collaborative setting is formulated around project-based or team-based scenarios~\cite{DBLP:conf/cikm/TolmachovaIX20}.
Research on understanding students online activity in SaL~\cite{jaakonmaki2020understanding} suggests that the majority of students primarily choose content-based web pages, like Wikipedia, and that visualisations are essential, both for students and teachers who later can better monitor students behaviour. This finding supports our idea of implementing PKGs and semantic technologies in the collaborative search and learning systems, and connecting the search with KGs and entities. 

Furthermore, different systems have been introduced in the literature for collaborative search. One of the newest advancements is the QueryTogether~\cite{DBLP:journals/ipm/AndolinaKRFJ18}, a system for entity-centric collaborative search in spontaneous settings that showed that entities could portray a vital role as interactive search objects. Also, their study confirmed that a common ground provided in collocated interaction supports group awareness, which has a positive impact on collaboration. Inspired by their work, we plan to emphasise the entities in the group search and implement features that aid group participation and awareness. 
Moreover, the Learnweb platform had implemented a semantic-based approach the LogCanvas~\cite{DBLP:conf/sigir/XuF0N18}, a visualisation tool linking to KGs for collaborative search. However, although the semantically enriched visual is a technically useful direction, the implementation lacks a collaborative and user-friendly interface and design. We will enhance this work by understanding the past pitfalls and designing collaborative semantic features linked to KGs.





%% file: 4-approach.tex
\section{Proposed approach}

\begin{figure}
    \centering
    \includegraphics[width=0.49\textwidth]{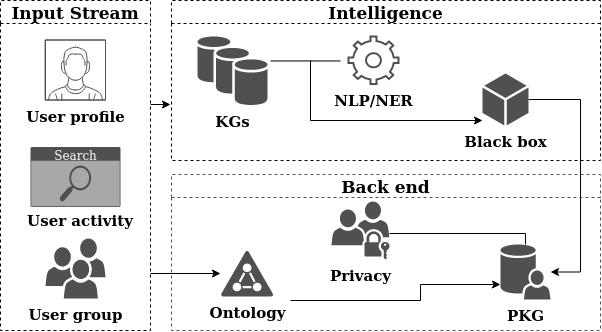}
    \caption{The overall architecture that includes the input stream, intelligence computation and back end creation of a personal knowledge graph (PKG). 
    }
    \Description[System architecture for creating PKG]{Input Stream, Intelligence, and back end parts of the system architecture for the creation of a personal knowledge graph. It is constructed by user profile, user activity, user group in the input stream; by knowledge graphs, named entity recognition and natural language processing software interconnected with the black box computation in the intelligence part; and by ontology and privacy to construct the personal knowledge graph in the back end.}
    \label{fig: backend}
\end{figure}

We propose an architecture for the creation of a PKG for a user in the back end of an e-learning system in Figure~\ref{fig: backend} (RQ1). The input stream consists of the user's generated data, which is then passed to the intelligence part. In combination with an ontology, the input data are processed with the named entity recognition (NER) and natural language processing (NLP) software to identify the concepts and entities from the KGs. Because user's actions change through time, there is a weighted algorithm that recalculates the main points of interest for the user in each period of time. The black box calculates those weights and offers a filter in the recognised entities from the KGs, which will be passed to the PKG. Then the PKG is created with respect to user's and user group privacy. 
By implementing the PKGs as a structural element of a platform, we interconnect the 
activities that happen in the platform with Linked Data and KGs. 
The research work in the creation of a PKG intends to address the questions related to syntax and semantics as structural elements and regarding the represented data in each application context (RQ1).


\subsection{Use case 1: Collaborative search}

Applications using web search can get benefited by the usage of PKGs, which can offer semantically enhanced features and personalisation capabilities. However, currently PKGs cannot play a role in the filter bubble of the search results provided to the users.
We develop the collaborativesemantically enhanc search use case in the 
e-learning platform Learnweb~\cite{Learnweb}. 
We link the input stream data with KGs to attain semantic relations between the data and identify the most important entities of the collaboration.

The proposed approach, in combination with a user-friendly interface and graphics, can help us reveal more personalised as well as group-project needs and develop collaborative features that are boosted with rich metadata and interconnections with the SW. Therefore, we would be able to examine how a collaborative e-learning platform can benefit from the addition of PKGs in their database to offer advanced SW features, personalisation (RQ3), and better collaboration in general (RQ4).
The potential applications we can implement with the usage of PKGs include but are not limited to better users' credibility and understanding of the group's activity, learning analytics, recommendations, and feedback.



\subsection{Use case 2: e-Learning}

In this use case, we utilise the knowledge base of the eDoer platform~\cite{eDoer}, an open learning recommendation system prototype that connects the labour market skills with open educational resources (OERs). The eDoer is currently focused on Data Science related skills; however, the research targets into system's integration of a knowledge base in general domain OERs. The implementation of our approach could allow users to receive personalised recommendations based on their learning preferences and needs, accessibility needs and access, and semantic-based solutions, such as relevant to their topics learning content (RQ2).

\subsection{Opportunities and Challenges}


Our proposed approach subsist on some novel items and opportunities. So far, the current problem has not been solved with the usage of PKGs. This research might increase the interest in integrating PKGs in other domains, as well as the influence of the usage of knowledge bases in the e-learning community to utilise publicly available KGs and PKGs in their applications.
An opportunity stands in the exploration of better collaborative features via the PKGs. Besides the plethora of collaborative search systems, no collaborative search interface has received major attention and has become long-established. This might happen due to the features offered in collaborative systems being adopted from single user purpose interfaces and are not directly developed for collaborative environments~\cite{DBLP:journals/computer/Hearst14}.





However, there are plenty of challenges this research needs to tackle. At first, we need to face the puzzle of knowledge acquisition, entity recognition and linking, storage and maintenance of the PKGs and connection to external PKGs. We select to start our alpha version by selecting DBpedia KG~\cite{DBLP:journals/semweb/LehmannIJJKMHMK15} and DBpedia Spotlight~\cite{DBLP:journals/ojsw/ChabchoubGZ18} to store our data locally in the Learnweb platform and maintain the data based on the processes provided by the Learnweb platform. In the future, we will focus on linking with more KGs and different NER software to improve our computation time and increase our coverage of search topics into more domains and languages. However, if the platform becomes widely used by thousands of users every day, we will be forced to expand our local implantation into a cloud service for computation and storage.

Also, by deploying PKGs, we need to raise privacy as a first-class citizen and ensure users' consent to select data from the personal profile, store user actions, and share part of them with the collaborators to assist the collaboration and offer richer features. Also, the right to be forgotten based on General Data Protection Regulation~\cite{voigt2017eu} should be provided to the single user as well as the group results. We target to face this obstacle by keeping the users' data only for a limited period of time, nonetheless long enough to serve the team's objectives. Another vital factor to consider is that PKGs are highly time dependant~\cite{DBLP:conf/ictir/BalogK19}. Their information might change, particularly depending on the computation time, which affects the applications offered based on PKGs back end.
Overall, we aim to address this part by setting time constrains on PKG computation time falls to ensure some level of confidence on the computed data. We discuss further challenges in future work.

%% file: 5-methodology.tex
\section{Methodology}

The methodology for approaching the research questions combines qualitative and quantitative analysis.
In order to understand how the SW technologies are used in the e-learning domain, we plan to conduct an interdisciplinary examination of the state-of-the-art solutions relevant to our studied problems. This investigation includes the literature in the user profiling ontologies and schemes, literature in PKGs, and publications of solutions that apply personalisation in KGs and e-learning.
Following, we aim to formalise our research questions and hypotheses and define potential solutions while identifying the innovative contributions. The formal characteristics of the proposed approach will be demonstrated, and the implemented solution will be made available to the community concerning users' privacy.

Mainly for evaluating our implementation, we will involve human participants in our research after receiving their consent based on national and international privacy laws. A similar work to ours~\cite{DBLP:conf/wsdm/SafaviFSJWFKB20}, evaluated their method with requiting some participants (10 people). People with some technical background could perform experiments in the platform, which will allow us to collect and process their data and have their own experiences of the applications, we will develop documented in the form of interviews and questionnaires, and perform qualitative analysis; 
it will measure 
users' experience and satisfaction of the implemented features.
Experts and professionals might also be included in our studies to receive precise feedback regarding the technical implementation, the learning aspects, and the user interface.

Nonetheless we are facing a limitation because there are no gold-standards and baseline metrics for evaluating collaborative search and SaL. Among the different metrics that have been proposed are the effects on users' short and long term memory and gained knowledge, the time it takes to find information online and others.
Several benchmarks have been suggested for a sub-domain of potential application tasks like the recommendations, such as the EdNet~\cite{DBLP:conf/aied/ChoiLSCPLBBKH20}; but, they do not fit the nature of SaL activities, project-based and team-based learning. However, we could compare our method with the state-of-the-art recommendation software by adopting them into a collaborative search learning environment.

%% file: 6-results.tex
\section{Results}\label{sec: results}



\begin{figure}
    \centering
    \includegraphics[width=0.5\textwidth]{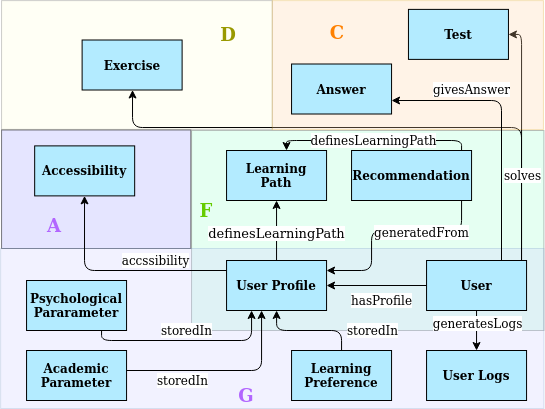}
    \caption{User Profile pattern that EduCOR~\cite{DBLP:conf/semweb/IlkouATHKAN21} ontology suggests.}
    \label{fig: patternG}
\end{figure}

\begin{figure}[h!]
    \centering
\begin{tikzpicture}

    \pgfplotsset{
        show sum on top/.style={
            /pgfplots/scatter/@post marker code/.append code={%
                \node[
                    at={(normalized axis cs:%
                            \pgfkeysvalueof{/data point/x},%
                            \pgfkeysvalueof{/data point/y})%
                    },
                    anchor=south,
                ]
                {\pgfmathprintnumber{\pgfkeysvalueof{/data point/y}}};
            },
        },
    }

  \begin{axis}[
    ybar stacked, ymin=0,  
    bar width=9mm,
    symbolic x coords={EQ1,EQ2,EQ3,EQ4,EQ5},
    xtick=data,
    nodes near coords, 
    legend style={at={(0.2,1.205)},anchor=west},
  ]
  \addplot [fill=blue] coordinates {
({EQ1},3.81)
({EQ2},1.91)
({EQ3},2.86)
({EQ4},3.81)
({EQ5},4.76)};
  \addplot [fill=red] coordinates {
({EQ1},14.29)
({EQ2},12.38)
({EQ3},9.52)
({EQ4},8.57)
({EQ5},7.62)
};
  \addplot [fill=yellow] coordinates {
({EQ1},21.90)
({EQ2},25.71)
({EQ3},18.09)
({EQ4},29.52)
({EQ5},19.05)
};
  \addplot [fill=green] coordinates {
({EQ1},36.19)
({EQ2},40.95)
({EQ3},39.05)
({EQ4},31.43)
({EQ5},33.33)
};
  \addplot [fill=black!40!green] coordinates {
({EQ1},23.81)
({EQ2},19.05)
({EQ3},30.48)
({EQ4},26.67)
({EQ5},35.24)
};
  \legend{Strongly disagree,Somewhat disagree, Neither agree or disagree, Somewhat agree, Strongly agree}
  \end{axis}
  \end{tikzpicture}
\caption{User general feedback from CollabGraph~\cite{collabgraph}. EQ1: I like the group results visualized in a graph, EQ2: I like the summary of the team-members results, EQ3: I like the graph visualizations , EQ4: I want to have a graph visualization  next to the list view of the search results, EQ5: I like the combination of the list and graph view.  }
    \label{fig: bar_chart_sum}
\end{figure}
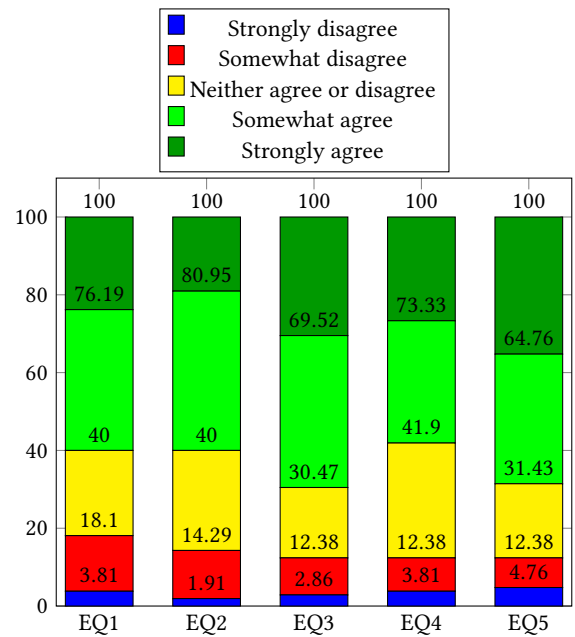

The current line of research is still in an early stage. 
In our first steps in user modelling in e-learning platforms, we have published EduCOR ontology~\cite{DBLP:conf/semweb/IlkouATHKAN21} for personalised recommendations of educational resources. EduCOR consists of different parts and has a focus on user profiling, as can be seen in Figure~\ref{fig: patternG}. 
This part of the overall ontology has been extended in our latest submission to facilitate the necessary components for the creation of PKG ontology for web search~\cite{pkgonto} with respect to accessibility parameters, such as content access rights and privacy.

In our latest paper, which is currently under submission~\cite{collabgraph}, we dive into the collaborative search and propose a collaborative search graph summary visualisation alongside the classical list-view of search results~\cite{collabgraph}. Our back end suggests the development of PKGs for each user which capture users' activity, connect to a KG in order to identify concepts and entities, and propagate the top entities to be visualised in the group summary graph. We evaluated our system on six different learning scenarios among 105 valid participants in a well-established user experience questionnaire and some evaluation questions we developed based on the parameters we wanted to rate. Our system shows high likability among users, as can be seen in Figure~\ref{fig: bar_chart_sum}. However, it is still to be investigated how users perceive our system in an on-site experiment, and how the implementation of PKGs is superior to a simpler approach with the extensions we foresee for our system.

%% file: 7-conclusion.tex
\section{Conclusions and future work}



We presented the research plan on personal knowledge graphs in collaborative search environments and e-learning platforms. We outlined our design based on the related work and discussed the methodology and proposed implementation in the two different use cases in e-learning domain. 
We argued that the proposed approach could benefit collaborative learning search systems and e-learning platforms which are connected to knowledge bases by connecting them to semantic technologies, and we suggested a few applications to be deployed. However, there are is a broad domain of future applications such systems can develop.



We believe this work opens a new line of research in web search and PKGs. 
At first, this research can explore the knowledge acquisition processes as well as the mainting, creation and update factors of PKGs.
Further, privacy is a constant concern of personalised features. This issue could be addressed with a collaboration with legal researchers. 
Moreover, editor's and author's data could align with the data offered in the KG and provide the next generation KGs which offer advanced content credibility, access rights and privacy. Additionally, this work could be further developed to offer semantic personalised recommendations. These could be related to further web search items in collaborative search setting or suggested topics and educational resources in e-learning platforms.

From a general viewpoint, this research could be deployed towards the broader e-learning field and the human factors in computing systems or human-computer interaction. One case could be the additional annotations and features to support learning, such as direct feedback from the teacher in highlighted text and comments, and more user-centric visualisations. Another suggestion from the educational perspective could be the investigation of PKGs outcomes in knowledge building spaces. 



%% file: main.bbl

\begin{thebibliography}{26}


\ifx \showCODEN    \undefined \def \showCODEN     #1{\unskip}     \fi
\ifx \showDOI      \undefined \def \showDOI       #1{#1}\fi
\ifx \showISBNx    \undefined \def \showISBNx     #1{\unskip}     \fi
\ifx \showISBNxiii \undefined \def \showISBNxiii  #1{\unskip}     \fi
\ifx \showISSN     \undefined \def \showISSN      #1{\unskip}     \fi
\ifx \showLCCN     \undefined \def \showLCCN      #1{\unskip}     \fi
\ifx \shownote     \undefined \def \shownote      #1{#1}          \fi
\ifx \showarticletitle \undefined \def \showarticletitle #1{#1}   \fi
\ifx \showURL      \undefined \def \showURL       {\relax}        \fi
\providecommand\bibfield[2]{#2}
\providecommand\bibinfo[2]{#2}
\providecommand\natexlab[1]{#1}
\providecommand\showeprint[2][]{arXiv:#2}

\bibitem[\protect\citeauthoryear{Abdi, Khosravi, Sadiq, and Gasevic}{Abdi
  et~al\mbox{.}}{2020}]%
        {abdi2020complementing}
\bibfield{author}{\bibinfo{person}{Solmaz Abdi}, \bibinfo{person}{Hassan
  Khosravi}, \bibinfo{person}{Shazia Sadiq}, {and} \bibinfo{person}{Dragan
  Gasevic}.} \bibinfo{year}{2020}\natexlab{}.
\newblock \showarticletitle{Complementing educational recommender systems with
  open learner models}. In \bibinfo{booktitle}{\emph{Proceedings of the tenth
  international conference on learning analytics \& knowledge}}.
\newblock


\bibitem[\protect\citeauthoryear{Andolina, Klouche, Ruotsalo, Flor{\'{e}}en,
  and Jacucci}{Andolina et~al\mbox{.}}{2018}]%
        {DBLP:journals/ipm/AndolinaKRFJ18}
\bibfield{author}{\bibinfo{person}{Salvatore Andolina}, \bibinfo{person}{Khalil
  Klouche}, \bibinfo{person}{Tuukka Ruotsalo}, \bibinfo{person}{Patrik
  Flor{\'{e}}en}, {and} \bibinfo{person}{Giulio Jacucci}.}
  \bibinfo{year}{2018}\natexlab{}.
\newblock \showarticletitle{Querytogether: Enabling entity-centric exploration
  in multi-device collaborative search}.
\newblock \bibinfo{journal}{\emph{Inf. Process. Manag.}}
  (\bibinfo{year}{2018}).
\newblock


\bibitem[\protect\citeauthoryear{Apoki}{Apoki}{2021}]%
        {apoki2021design}
\bibfield{author}{\bibinfo{person}{Ufuoma~Chima Apoki}.}
  \bibinfo{year}{2021}\natexlab{}.
\newblock \showarticletitle{The design of WASPEC: A fully personalised Moodle
  system using semantic web technologies}.
\newblock \bibinfo{journal}{\emph{Computers}} (\bibinfo{year}{2021}).
\newblock


\bibitem[\protect\citeauthoryear{Balog and Kenter}{Balog and Kenter}{2019}]%
        {DBLP:conf/ictir/BalogK19}
\bibfield{author}{\bibinfo{person}{Krisztian Balog} {and} \bibinfo{person}{Tom
  Kenter}.} \bibinfo{year}{2019}\natexlab{}.
\newblock \showarticletitle{Personal Knowledge Graphs: {A} Research Agenda}. In
  \bibinfo{booktitle}{\emph{Proceedings of the 2019 {ACM} {SIGIR} International
  Conference on Theory of Information Retrieval, {ICTIR} 2019, Santa Clara, CA,
  USA, October 2-5, 2019}}.
\newblock


\bibitem[\protect\citeauthoryear{C. and Reddy}{C. and Reddy}{2019}]%
        {DBLP:journals/jms/CR19}
\bibfield{author}{\bibinfo{person}{Senthilkumar~N. C.} {and}
  \bibinfo{person}{Ch.~Pradeep Reddy}.} \bibinfo{year}{2019}\natexlab{}.
\newblock \showarticletitle{Collaborative Search Engine for Enhancing
  Personalized User Search Based on Domain Knowledge}.
\newblock \bibinfo{journal}{\emph{J. Medical Syst.}} (\bibinfo{year}{2019}).
\newblock


\bibitem[\protect\citeauthoryear{Chabchoub, Gagnon, and Zouaq}{Chabchoub
  et~al\mbox{.}}{2018}]%
        {DBLP:journals/ojsw/ChabchoubGZ18}
\bibfield{author}{\bibinfo{person}{Mohamed Chabchoub}, \bibinfo{person}{Michel
  Gagnon}, {and} \bibinfo{person}{Amal Zouaq}.}
  \bibinfo{year}{2018}\natexlab{}.
\newblock \showarticletitle{{FICLONE:} Improving DBpedia Spotlight Using Named
  Entity Recognition and Collective Disambiguation}.
\newblock \bibinfo{journal}{\emph{Open J. Semantic Web}}
  (\bibinfo{year}{2018}).
\newblock


\bibitem[\protect\citeauthoryear{Choi, Lee, Shin, Cho, Park, Lee, Baek, Bae,
  Kim, and Heo}{Choi et~al\mbox{.}}{2020}]%
        {DBLP:conf/aied/ChoiLSCPLBBKH20}
\bibfield{author}{\bibinfo{person}{Youngduck Choi}, \bibinfo{person}{Youngnam
  Lee}, \bibinfo{person}{Dongmin Shin}, \bibinfo{person}{Junghyun Cho},
  \bibinfo{person}{Seoyon Park}, \bibinfo{person}{Seewoo Lee},
  \bibinfo{person}{Jineon Baek}, \bibinfo{person}{Chan Bae},
  \bibinfo{person}{Byungsoo Kim}, {and} \bibinfo{person}{Jaewe Heo}.}
  \bibinfo{year}{2020}\natexlab{}.
\newblock \showarticletitle{EdNet: {A} Large-Scale Hierarchical Dataset in
  Education}. In \bibinfo{booktitle}{\emph{Artificial Intelligence in Education
  - 21st International Conference, {AIED} 2020, Ifrane, Morocco, July 6-10,
  2020, Proceedings, Part {II}}} \emph{(\bibinfo{series}{Lecture Notes in
  Computer Science})}. \bibinfo{publisher}{Springer}.
\newblock


\bibitem[\protect\citeauthoryear{Davies, Lehdonvirta, Margaryan, Albert, and
  Larke}{Davies et~al\mbox{.}}{2020}]%
        {davies2020developing}
\bibfield{author}{\bibinfo{person}{H Davies}, \bibinfo{person}{V Lehdonvirta},
  \bibinfo{person}{A Margaryan}, \bibinfo{person}{J Albert}, {and}
  \bibinfo{person}{LR Larke}.} \bibinfo{year}{2020}\natexlab{}.
\newblock \showarticletitle{Developing and matching skills in the online
  platform economy: Findings on new forms of digital work and learning from
  Cedefop’s CrowdLearn study}.
\newblock  (\bibinfo{year}{2020}).
\newblock


\bibitem[\protect\citeauthoryear{Faber, Safavi, Mottin, M{\"u}ller, and
  Koutra}{Faber et~al\mbox{.}}{2018}]%
        {faber2018adaptive}
\bibfield{author}{\bibinfo{person}{Lukas Faber}, \bibinfo{person}{Tara Safavi},
  \bibinfo{person}{Davide Mottin}, \bibinfo{person}{Emmanuel M{\"u}ller}, {and}
  \bibinfo{person}{Danai Koutra}.} \bibinfo{year}{2018}\natexlab{}.
\newblock \showarticletitle{Adaptive Personalized Knowledge Graph
  Summarization}. In \bibinfo{booktitle}{\emph{MLG Workshop (with KDD)}}.
\newblock


\bibitem[\protect\citeauthoryear{Hearst}{Hearst}{2014}]%
        {DBLP:journals/computer/Hearst14}
\bibfield{author}{\bibinfo{person}{Marti~A. Hearst}.}
  \bibinfo{year}{2014}\natexlab{}.
\newblock \showarticletitle{What's Missing from Collaborative Search?}
\newblock \bibinfo{journal}{\emph{Computer}} (\bibinfo{year}{2014}).
\newblock


\bibitem[\protect\citeauthoryear{Ilkou, Abu{-}Rasheed, Tavakoli, Hakimov,
  Kismih{\'{o}}k, Auer, and Nejdl}{Ilkou et~al\mbox{.}}{2021}]%
        {DBLP:conf/semweb/IlkouATHKAN21}
\bibfield{author}{\bibinfo{person}{Eleni Ilkou}, \bibinfo{person}{Hasan
  Abu{-}Rasheed}, \bibinfo{person}{MohammadReza Tavakoli},
  \bibinfo{person}{Sherzod Hakimov}, \bibinfo{person}{G{\'{a}}bor
  Kismih{\'{o}}k}, \bibinfo{person}{S{\"{o}}ren Auer}, {and}
  \bibinfo{person}{Wolfgang Nejdl}.} \bibinfo{year}{2021}\natexlab{}.
\newblock \showarticletitle{EduCOR: An Educational and Career-Oriented
  Recommendation Ontology}. In \bibinfo{booktitle}{\emph{The Semantic Web -
  {ISWC} 2021 - 20th International Semantic Web Conference, {ISWC} 2021,
  Virtual Event, October 24-28, 2021, Proceedings}}
  \emph{(\bibinfo{series}{Lecture Notes in Computer Science})}.
  \bibinfo{publisher}{Springer}.
\newblock


\bibitem[\protect\citeauthoryear{Ilkou and Signer}{Ilkou and Signer}{2020}]%
        {ilkou2020technology}
\bibfield{author}{\bibinfo{person}{Eleni Ilkou} {and} \bibinfo{person}{Beat
  Signer}.} \bibinfo{year}{2020}\natexlab{}.
\newblock \showarticletitle{A Technology-enhanced Smart Learning Environment
  based on the Combination of Knowledge Graphs and Learning Paths.}. In
  \bibinfo{booktitle}{\emph{CSEDU (2)}}. \bibinfo{pages}{461--468}.
\newblock


\bibitem[\protect\citeauthoryear{Ilkou, Taibi, Fisichella, and
  Tolmachova}{Ilkou et~al\mbox{.}}{2022a}]%
        {pkgonto}
\bibfield{author}{\bibinfo{person}{Eleni Ilkou}, \bibinfo{person}{Davide
  Taibi}, \bibinfo{person}{Marco Fisichella}, {and} \bibinfo{person}{Tetiana
  Tolmachova}.} \bibinfo{year}{2022}\natexlab{a}.
\newblock \bibinfo{title}{Personal Knowledge Graph Ontology for Web Search}.
  (\bibinfo{year}{2022}).
\newblock


\bibitem[\protect\citeauthoryear{Ilkou, Tolmachova, Fisichella, and
  Taibi}{Ilkou et~al\mbox{.}}{2022b}]%
        {collabgraph}
\bibfield{author}{\bibinfo{person}{Eleni Ilkou}, \bibinfo{person}{Tetiana
  Tolmachova}, \bibinfo{person}{Marco Fisichella}, {and}
  \bibinfo{person}{Davide Taibi}.} \bibinfo{year}{2022}\natexlab{b}.
\newblock \bibinfo{title}{CollabGraph: A graph-based collaborative search
  summary visualisation}.  (\bibinfo{year}{2022}).
\newblock


\bibitem[\protect\citeauthoryear{Ismail, Belkhouche, and Harous}{Ismail
  et~al\mbox{.}}{2019}]%
        {ismail2019framework}
\bibfield{author}{\bibinfo{person}{Heba~M Ismail}, \bibinfo{person}{Boumediene
  Belkhouche}, {and} \bibinfo{person}{Saad Harous}.}
  \bibinfo{year}{2019}\natexlab{}.
\newblock \showarticletitle{Framework for personalized content recommendations
  to support informal learning in massively diverse information Wikis}.
\newblock \bibinfo{journal}{\emph{IEEE Access}} (\bibinfo{year}{2019}).
\newblock


\bibitem[\protect\citeauthoryear{Jaakonm{\"a}ki, vom Brocke, Dietze, Drachsler,
  Fortenbacher, Helbig, Kickmeier-Rust, Marenzi, Suarez, and
  Yun}{Jaakonm{\"a}ki et~al\mbox{.}}{2020}]%
        {jaakonmaki2020understanding}
\bibfield{author}{\bibinfo{person}{Roope Jaakonm{\"a}ki}, \bibinfo{person}{Jan
  vom Brocke}, \bibinfo{person}{Stefan Dietze}, \bibinfo{person}{Hendrik
  Drachsler}, \bibinfo{person}{Albrecht Fortenbacher},
  \bibinfo{person}{Ren{\'e} Helbig}, \bibinfo{person}{Michael Kickmeier-Rust},
  \bibinfo{person}{Ivana Marenzi}, \bibinfo{person}{Angel Suarez}, {and}
  \bibinfo{person}{Haeseon Yun}.} \bibinfo{year}{2020}\natexlab{}.
\newblock \showarticletitle{Understanding Students’ Online Behavior While
  They Search on the Internet: Searching as Learning}.
\newblock In \bibinfo{booktitle}{\emph{Learning Analytics Cookbook}}.
\newblock


\bibitem[\protect\citeauthoryear{Jones}{Jones}{2007}]%
        {DBLP:journals/arist/Jones07}
\bibfield{author}{\bibinfo{person}{William Jones}.}
  \bibinfo{year}{2007}\natexlab{}.
\newblock \showarticletitle{Personal Information Management}.
\newblock \bibinfo{journal}{\emph{Annu. Rev. Inf. Sci. Technol.}}
  (\bibinfo{year}{2007}).
\newblock


\bibitem[\protect\citeauthoryear{Labs}{Labs}{2021}]%
        {eDoer}
\bibfield{author}{\bibinfo{person}{TIB Labs}.} \bibinfo{year}{2021}\natexlab{}.
\newblock \bibinfo{booktitle}{\emph{{eDoer} Education Portal}}.
\newblock
\urldef\tempurl%
\url{https://labs.tib.eu/edoer/}
\showURL{%
\tempurl}


\bibitem[\protect\citeauthoryear{Learnweb}{Learnweb}{2021}]%
        {Learnweb}
\bibfield{author}{\bibinfo{person}{Learnweb}.} \bibinfo{year}{2021}\natexlab{}.
\newblock \bibinfo{booktitle}{\emph{{Learnweb} - Learnweb}}.
\newblock
\urldef\tempurl%
\url{https://learnweb.l3s.uni-hannover.de/}
\showURL{%
\tempurl}


\bibitem[\protect\citeauthoryear{Lehmann, Isele, Jakob, Jentzsch, Kontokostas,
  Mendes, Hellmann, Morsey, van Kleef, Auer, and Bizer}{Lehmann
  et~al\mbox{.}}{2015}]%
        {DBLP:journals/semweb/LehmannIJJKMHMK15}
\bibfield{author}{\bibinfo{person}{Jens Lehmann}, \bibinfo{person}{Robert
  Isele}, \bibinfo{person}{Max Jakob}, \bibinfo{person}{Anja Jentzsch},
  \bibinfo{person}{Dimitris Kontokostas}, \bibinfo{person}{Pablo~N. Mendes},
  \bibinfo{person}{Sebastian Hellmann}, \bibinfo{person}{Mohamed Morsey},
  \bibinfo{person}{Patrick van Kleef}, \bibinfo{person}{S{\"{o}}ren Auer},
  {and} \bibinfo{person}{Christian Bizer}.} \bibinfo{year}{2015}\natexlab{}.
\newblock \showarticletitle{DBpedia - {A} large-scale, multilingual knowledge
  base extracted from Wikipedia}.
\newblock \bibinfo{journal}{\emph{Semantic Web}} (\bibinfo{year}{2015}).
\newblock


\bibitem[\protect\citeauthoryear{Safavi, Belth, Faber, Mottin, M{\"{u}}ller,
  and Koutra}{Safavi et~al\mbox{.}}{2019}]%
        {DBLP:conf/icdm/SafaviBFMMK19}
\bibfield{author}{\bibinfo{person}{Tara Safavi}, \bibinfo{person}{Caleb Belth},
  \bibinfo{person}{Lukas Faber}, \bibinfo{person}{Davide Mottin},
  \bibinfo{person}{Emmanuel M{\"{u}}ller}, {and} \bibinfo{person}{Danai
  Koutra}.} \bibinfo{year}{2019}\natexlab{}.
\newblock \showarticletitle{Personalized Knowledge Graph Summarization: From
  the Cloud to Your Pocket}. In \bibinfo{booktitle}{\emph{2019 {IEEE}
  International Conference on Data Mining, {ICDM} 2019, Beijing, China,
  November 8-11, 2019}}. \bibinfo{publisher}{{IEEE}}.
\newblock


\bibitem[\protect\citeauthoryear{Safavi, Fourney, Sim, Juraszek, Williams,
  Friend, Koutra, and Bennett}{Safavi et~al\mbox{.}}{2020}]%
        {DBLP:conf/wsdm/SafaviFSJWFKB20}
\bibfield{author}{\bibinfo{person}{Tara Safavi}, \bibinfo{person}{Adam
  Fourney}, \bibinfo{person}{Robert Sim}, \bibinfo{person}{Marcin Juraszek},
  \bibinfo{person}{Shane Williams}, \bibinfo{person}{Ned Friend},
  \bibinfo{person}{Danai Koutra}, {and} \bibinfo{person}{Paul~N. Bennett}.}
  \bibinfo{year}{2020}\natexlab{}.
\newblock \showarticletitle{Toward Activity Discovery in the Personal Web}. In
  \bibinfo{booktitle}{\emph{{WSDM} '20: The Thirteenth {ACM} International
  Conference on Web Search and Data Mining, Houston, TX, USA, February 3-7,
  2020}}. \bibinfo{publisher}{{ACM}}.
\newblock


\bibitem[\protect\citeauthoryear{Tolmachova, Ilkou, and Xu}{Tolmachova
  et~al\mbox{.}}{2020}]%
        {DBLP:conf/cikm/TolmachovaIX20}
\bibfield{author}{\bibinfo{person}{Tetiana Tolmachova}, \bibinfo{person}{Eleni
  Ilkou}, {and} \bibinfo{person}{Luyan Xu}.} \bibinfo{year}{2020}\natexlab{}.
\newblock \showarticletitle{Working Towards the Ideal Search History
  Interface}. In \bibinfo{booktitle}{\emph{Proceedings of the {CIKM} 2020
  Workshops co-located with 29th {ACM} International Conference on Information
  and Knowledge Management {(CIKM} 2020), Galway, Ireland, October 19-23,
  2020}}. \bibinfo{publisher}{CEUR-WS.org}.
\newblock


\bibitem[\protect\citeauthoryear{Voigt and Von~dem Bussche}{Voigt and Von~dem
  Bussche}{2017}]%
        {voigt2017eu}
\bibfield{author}{\bibinfo{person}{Paul Voigt} {and} \bibinfo{person}{Axel
  Von~dem Bussche}.} \bibinfo{year}{2017}\natexlab{}.
\newblock \showarticletitle{The eu general data protection regulation (gdpr)}.
\newblock \bibinfo{journal}{\emph{A Practical Guide, 1st Ed., Cham: Springer
  International Publishing}} (\bibinfo{year}{2017}).
\newblock


\bibitem[\protect\citeauthoryear{Wang, Miao, and Yang}{Wang
  et~al\mbox{.}}{2018}]%
        {wang2018design}
\bibfield{author}{\bibinfo{person}{Huaqiong Wang}, \bibinfo{person}{Xiaoyu
  Miao}, {and} \bibinfo{person}{Pan Yang}.} \bibinfo{year}{2018}\natexlab{}.
\newblock \showarticletitle{Design and implementation of personal health record
  systems based on knowledge graph}. In \bibinfo{booktitle}{\emph{2018 9th
  International Conference on Information Technology in Medicine and Education
  (ITME)}}.
\newblock


\bibitem[\protect\citeauthoryear{Xu, Fernando, Zhou, and Nejdl}{Xu
  et~al\mbox{.}}{2018}]%
        {DBLP:conf/sigir/XuF0N18}
\bibfield{author}{\bibinfo{person}{Luyan Xu}, \bibinfo{person}{Zeon~Trevor
  Fernando}, \bibinfo{person}{Xuan Zhou}, {and} \bibinfo{person}{Wolfgang
  Nejdl}.} \bibinfo{year}{2018}\natexlab{}.
\newblock \showarticletitle{LogCanvas: Visualizing Search History Using
  Knowledge Graphs}. In \bibinfo{booktitle}{\emph{The 41st International {ACM}
  {SIGIR} Conference on Research {\&} Development in Information Retrieval,
  {SIGIR} 2018, Ann Arbor, MI, USA, July 08-12, 2018}}.
  \bibinfo{publisher}{{ACM}}.
\newblock


\end{thebibliography}
